\documentclass[aps,pre,preprint,groupedaddress,showpacs]{revtex4-2}

\usepackage{graphicx}
\usepackage{amssymb}
\usepackage{color} 
\usepackage{epstopdf}
\usepackage{soul}
\newcommand{\corr}[1]{\textcolor{black}{#1}} 
\newcommand{\corrnew}[1]{\textcolor{black}{#1}} 

\begin{document}


\title{Undulatory forcing of an intruder through granular media: effects of frequency and packing fraction\\
	\textnormal{Accepted manuscript for Physical Review E, 113, 055422 (2026), DOI: 10.1103/xb4r-bcxp, https://doi.org/10.1103/xb4r-bcxp}}



\author{Douglas D. Carvalho}
\author{Erick M. Franklin}
 \email{erick.franklin@unicamp.br}
 \thanks{Corresponding author}
\affiliation{%
Faculdade de Engenharia Mec\^anica, Universidade Estadual de Campinas (UNICAMP)\\
Rua Mendeleyev, 200, Campinas, SP, Brazil\\
}%


\date{\today}

\begin{abstract}
We investigate the motion amid grains of an intruder undergoing an imposed force that oscillates with a given frequency. For that, we made use of discrete numerical simulations where the intruder was a larger disk on which a force oscillating in direction was applied, and the grains consisted of smaller disks. All disks were placed on a surface with basal friction over which they could slide, the system was confined in the sliding directions, and we varied the system packing fraction, oscillation frequency, and magnitude of the forcing. The results show intermittent and very complex motions of the intruder depending on both the packing fraction and frequency of oscillation: it can move sideways while slowly progressing forward, it can be blocked during a long period after and/or before start moving, or it can simply be blocked after a given time. Interestingly, we find that the displacement velocity is much higher when the system packing fraction is above a given threshold, contrary to intuition. The results show that there is an optimal frequency that minimizes the transit time for some ranges of packing fraction, and we propose a model based on the system elasticity that explains this behavior and agrees with the numerical simulations. Our findings shed new light on how to better explore oscillating motion to move objects within granular media.

\end{abstract}


\maketitle


\section{\label{sec:Intro} INTRODUCTION}

Moving a solid object (intruder) in a granular medium is not as simple a task as in a fluid. If a moderate force is imposed, the object will present an intermittent motion that depends on solid-solid contacts: the formation of contact chains can be strong enough to deviate the intruder sideways or even to stop it, while their breakage allows motion in the direction of force application to resume (if a moderate velocity is imposed, this appears as strong fluctuations in the drag force) \cite{Kolb1, Tordesillas, Seguin1, Kozlowski, Carlevaro, Carvalho}. In addition, if the space is confined and the concentration of grains is high the system can jam \cite{Cates, Majmudar, Bi, Seguin2, Behringer_1}. On the other extreme, if the applied force is high and particle concentration small, high velocity fluctuations appear and particle-particle collisions play an important role in resisting the intruder's motion \cite{Bester}.

The low-velocity case is an interesting one, which appears, for example, in the locomotion of lizards, scorpions, worms, snakes and other animals in sand, the growth of plant roots in the soil, and also the penetration of probes in granular terrains. Those cases are characterized by a motion dominated by friction in solid-solid contacts, known as \textit{quasistatic regime} \cite{Andreotti_6}. \corr{When granular materials evolve in dense and quasistatic regimes, inter-particle forces are transmitted through a history-dependent contact network, often leading to anisotropic stress distributions \cite{Radjai1, Majmudar}. Local reorganizations of the packing \cite{Houssais} give rise to intermittent jamming and unjamming regions, depending on the persistence or failure of contact networks \cite{Cates, Majmudar, Bi, Seguin2, Behringer_1}, such that grains may either sustain strong external loads or yield under weaker forcing. In intruder problems, the continuous formation and rupture of these force networks results in a fluctuating drag force \cite{Kolb1, Seguin1}, which increases as the packing fraction approaches the jamming threshold. Over the past decades, considerable effort has been devoted to understanding stress transmission and jamming under both normal and shear loading \cite{Radjai1, Cates, Majmudar, Bi, Seguin2}. In particular, Radjai et al. \cite{Radjai1} showed that stress transmission in two-dimensional packings under biaxial compression occurs through two complementary networks: a load-bearing network of strong, nonsliding contacts that carries the deviatoric stress, and a dissipative network of weak, sliding contacts contributing primarily to the isotropic pressure. Along similar lines, Cates et al. \cite{Cates} introduced the concept of fragile states, in which force chains align along preferential directions, enabling the material to support loads anisotropically, while Bi et al. \cite{Bi} further showed that shear can induce both fragile states and shear jamming at packing fractions below those required for isotropic jamming. More recently, Carvalho et al. \cite{Carvalho} showed the existence of two analogous force networks when intruders move within a confined granular system: a bearing network that percolates forces from the intruder to the system boundaries, giving rise to jammed regions and large drag forces, and a dissipative network transmitting weaker forces within the bulk. They further showed that grains within the bearing chains undergo creep as the chains progressively break, elucidating the mechanism by which these structures collapse and allow the intruder to advance.}

\corr{Therefore, in the quasistatic regime} it is not uncommon to have a packed medium and reach jammed states \cite{Kolb1, Seguin1}, so that animals, plants and humans need to search for optimal ways of displacement and breaking jam \cite{Biewener}. One strategy adopted by some animals moving in sand is to propagate waves along their body in order to engage in succession the solid- and fluid-like behaviors of the granular medium \cite{Baumgartner, Maladen, Shimada}. This capacity led to some researches on the mechanics of those motions. For instance, Baumgartner et al. \cite{Baumgartner} and Maladen et al. \cite{Maladen} investigated the subsurface motion of the sandfish (\textit{Scincus scincus}), a lizard that can move long distances within sand to escape heat and predators, and can be found in North Africa and the Arabian peninsula \cite{news_science}. Although Baumgartner et al. \cite{Baumgartner} proposed that the sandfish propels itself using its limbs (based on results from nuclear magnetic resonance), Maladen et al. \cite{Maladen} showed that the limbs remain close to the sandfish body while it moves below the sand surface (measurements using high-speed x-ray imaging). Maladen et al. \cite{Maladen} showed that, in fact, the sandfish can move at speeds up to 10 cm/s by bending its body in an undulatory way. The oscillations correspond to high-amplitude transverse waves, with an efficiency that is independent of the packing fraction $\phi$ of the medium, the propelling velocity being controlled by the frequency of oscillation. The authors developed a model based on Resistive Force Theory (RFT) in which the intruder was split into different moving elements, and could estimate the wave efficiency and find the optimal kinematics. Those results were \corr{later} corroborated by Maladen et al. \cite{Maladen2} and Ding et al. \cite{Ding} by using Discrete Element Method (DEM) computations and also a sandfish-inspired robot, with a tendency of RFT to overestimate thrust. The RFT model of Maladen et al. \cite{Maladen} was later employed by Peng et al. \cite{Peng} to investigate finite and infinite slender intruders moving due to undulatory motion, and they showed that there are similarities between motion in granular media and viscous fluids, with the sawtooth being the optimal waveform \cite{Lighthill} for propulsion speed at a given power consumption.

Shimada et al. \cite{Shimada} investigated numerically the motion of lizards by using longitudinal waves, i.e., by alternately expanding its hindquarters and thrusting its contracted forequarters and head, and then pulling back the contracted hindquarters while expanding the forequarters. In their event-driven simulations, the intruder was modeled as disks that were connected with a liner spring, and that could alternately inflate and shrink as their distance increased or decreased. The authors found an optimal frequency for motion, given by the characteristic times of fluidizing sand around the moving part while anchoring sand around the inflated portion. More recently, Echeverr\'ia et al. \cite{Echeverria} investigated numerically the displacement of a long intruder oscillating transversely amid smaller disks (granular medium), all of them placed over a surface. In their simulations, the intruder consisted of a string of seven disks linked by damped springs, and all disks had a basal friction with the surface. The basal friction was varied in order to emulate different depths in real three-dimensional (3D) granular media. By varying the oscillation frequency, Echeverr\'ia et al. \cite{Echeverria} found a non-monotonic dependence of the displacement velocity with the frequency, and that the second vibrational mode of the intruder optimizes its displacement. The latter is explained in terms of number of contacts aligned with the direction of motion, which is higher for the second mode independent of the size of the system or those of grains.

Although considerable progress has been made in understanding the effect of oscillation in the displacement of intruders, there are some fundamental questions that still remain to be answered: (i) how the entire system responds to the oscillation frequencies? (ii) is there an optimal frequency for displacing a forced intruder in the desired direction? (iii) If yes, how it depends on the system properties (types of grains and packing fraction)? These are the questions that we intend to answer in this paper. For that, we carried out  Discrete Element Method (DEM) computations where the granular matter was an assembly of bidisperse disks and the intruder was a larger disk, all of them placed on a plane with basal friction (both dynamic and static) over which they could slide, in addition to the friction on solid-solid contacts (of their sides). A force whose direction oscillated in the plane of sliding was imposed on the intruder, and we varied the system packing fraction, and oscillation frequency and magnitude of the forcing. \corr{We impose a sinusoidal oscillation since some animals oscillate in a roughly sinusoidal form \cite{Maladen, Maladen2}, and, in addition, it is a simple waveform whose effects can be understood easier than other forms. The use of disks is also for keeping the system simple, allowing us to understand some key aspects of the undulatory motion.} The results show intermittent and very complex motions of the intruder depending on both the packing fraction and frequency of oscillation: it can move sideways while slowly progressing forward, it can be blocked during a long period after and/or before start moving, or it can simply be blocked after a given time. The results show that there is an optimal frequency that minimizes the transit time for some ranges of packing fraction, and we propose a model based on the system elasticity that shows that the optimal frequency corresponds approximately to the natural frequency of the entire system. Our results can be further explored for finding optimal methods of perforation or locomotion within sand and other grains. 

In the following, Secs. \ref{sec:model} and \ref{sec:setup} present, respectively, the model equations and numerical setup, and Sec. \ref{sec:Res} presents the results for the motion under different packing fractions and frequencies of forcing. Section \ref{sec:Conclu} presents the conclusions.

\section{\label{sec:model} MODEL EQUATIONS}

The numerical simulations were carried out with {DEM} \cite{Cundall}, in which the granular matter was an assembly of bidisperse disks and the intruder was a larger disk, all of them placed on a plane with basal friction. For the DEM computations, we made use of the open-source code LIGGGHTS \cite{Kloss, Berger}, and for producing disks with friction between particles and with the lateral walls we used the DESIgn toolbox \cite{Herman}. Finally, basal friction (both static and dynamic coefficients) was implemented in Carvalho et al. \cite{Carvalho}.

The motion of each particle, including the intruder, is governed by the linear (Eq. \ref{Fp}) and angular (Eq. \ref{Tp}) momentum equations,

\begin{equation}
	m\frac{d\vec{u}}{dt}= \vec{F}_{h} + \vec{F}_{c} + m\vec{g}\,\,,
	\label{Fp}
\end{equation}

\begin{equation}
	I\frac{d\vec{\omega}}{dt}=\vec{T}_{c} \,\,,
	\label{Tp}
\end{equation}

\noindent where $\vec{g}$ is the acceleration of gravity and, for each particle, $m$ is the mass, $\vec{u}$ is the velocity, $I$ is the moment of inertia, $\vec{\omega}$ is the angular velocity, $\vec{F}_{c}$ is the resultant of contact forces between solids, and $\vec{T}_{c}$ is the resultant of contact torques between solids. $\vec{F}_{h}$ is the oscillating forcing (described next) imposed on the intruder (exists only for the intruder). The contact forces are computed by Eq. \ref{Fc},

\begin{equation}
	\vec{F}_{c} = \sum_{i\neq j}^{N_c} \left(\vec{F}_{c,ij} \right) + \sum_{i}^{N_w} \left( \vec{F}_{c,iw} \right) \,\,,
	\label{Fc}
\end{equation}

\noindent and the contact torques are computed by Eq. \ref{Tc},

\begin{equation}
	\vec{T}_{c} = \sum_{i\neq j}^{N_c} \vec{T}_{c,ij} + \sum_{i}^{N_w} \vec{T}_{c,iw} \,\,,
	\label{Tc}
\end{equation}

\noindent where $\vec{F}_{c,ij}$ and $\vec{F}_{c,iw}$ are the contact forces between particles $i$ and $j$ and between particle $i$ and the wall, respectively, $\vec{T}_{c,ij}$ is the torque due to the tangential component of the contact force between particles $i$ and $j$, and $\vec{T}_{c,iw}$ is the torque due to the tangential component of the contact force between particle $i$ and the vertical wall. $N_c$ - 1 is the number of particles in contact with particle $i$, and $N_w$ the number of \corr{contacts with} the wall. For the contact forces between particles ($\vec{F}_{c,ij}$), and between particles and the lateral walls ($\vec{F}_{c,iw}$), the elastic Hertz-Mindlin contact model \cite{direnzo} is used (combination of two spring-dashpots, one for the normal and the other for the tangential interactions).

As in Carvalho et al. \cite{Carvalho}, $\vec{F}_{c,iw}$ includes the friction force between the bottom wall and each grain (basal friction), which follows the Coulomb law with static ($\mu_{s,g}$) and dynamic ($\mu_{d,g}$) coefficients: for a grain $i$ moving at a speed $v_{i} = |\vec{u}_{i}|$ above a threshold value $v^{\prime}$, a dynamic friction force with the bottom wall $\vec{F}_{c,iw}$ = $-\mu_{d,g}m_{i}|\vec{g}|\vec{u}_{i}/|\vec{u}_{i}|$ is considered, otherwise (if $v_{i} \leq v^{\prime}$)  a static friction force with the bottom wall $\vec{F}_{c,iw}$ = $-\mu_{s,g}m_{i}|\vec{g}|\vec{u}_{i}/|\vec{u}_{i}|$ is applied and the particle is stopped (by imposing $v_{i}$ = 0).

In order to thrust the intruder, we apply a harmonic-oscillating force $\vec{F}_h(t)$ on the intruder, given by Eq. \ref{eqHarmonicforce},

\begin{equation}
	\vec{F}_h(t) = |F_0 \cos(\omega t)| \hat{i} + F_0 \sin(\omega t) \hat{j} \,\,,
	\label{eqHarmonicforce}
\end{equation}

\noindent where $F_0$ is the force amplitude, $\omega$ is the angular frequency, $t$ is time, and $\hat{i}$ and $\hat{j}$ are unit vectors in the longitudinal and transverse directions \corr{(with respect to the domain)}, respectively.

\section{\label{sec:setup} NUMERICAL SETUP}

The setup is similar to that of Carvalho et al. \cite{Carvalho}, consisting of a monolayer of disks placed on a horizontal surface (in the $xy$ plane, $z$ being the coordinate aligned with gravity) over which they can slide, and confined by vertical walls. The granular matter consisted of disks with the mechanical properties of polyurethane (PSM-4), and with small and large diameters of $d_{s} = 4$ mm and $d_{l} = 5$ mm, respectively (in order to prevent crystallization \cite{Speedy}), and height $h_{g} = 3.2$ mm. The intruder had the mechanical properties of steel, with diameter and height of $d_{int} = 16$ mm and $h_{int} = 3.6$ mm, respectively. For the horizontal surface and vertical walls, we used the mechanical properties of glass. The system dimensions were of  $L_x$ $\times$ $L_{y}$ = 400 mm $\times$ 400  mm, where $L_x$ and $L_y$ are the longitudinal and transverse lengths, respectively, corresponding to approximately 35 $d_g$ $\times$ 35 $d_g$, where $d_g$ = 4.5 mm is the
mean diameter of disks. The number of large $N_{l}$ and small $N_{s}$ disks were set respecting a proportion $N_{l}/N_{s} \approx 0.64$ in all our simulations. Before each simulation, the grains were initially placed at random positions within a region exceeding the final computational domain. These grains were then gradually compressed toward the center of the domain until the target area was fully occupied. This approach allowed for efficient initialization of the desired packing fraction, as directly placing grains at high densities would be computationally prohibitive (in terms of time). Following the compaction stage, the system was allowed to relax until a low level of kinetic energy of the grains was achieved \corr{(because of dissipation due to the Hertz-Mindlin model and basal friction)}. All simulations were performed using a time step of $\Delta t$ $=$ 3.2 $\times$ 10$^{-6}$ s, which remained below 10$\%$ of the Rayleigh time in every case considered \cite{Derakhshani}.

The total domain was fixed in all simulations, so that the system packing fraction was controlled by varying the number of disks in each simulation, being varied within 0.760 $\leq$ $\phi$ $\leq$ 0.810 (the number of small and large disks used in the simulations are shown in Tab. \ref{tab_number_grains}). We used values found in the literature \cite{Carlevaro, Hashemnia, Gondret, Zaikin} for the Young's modulus $E$, Poisson ratio $\nu$, and coefficients of restitution $\epsilon$ and friction $\mu$ (static and dynamic). The friction between disks (including the intruder) with the bottom wall was implemented in Carvalho et al. \cite{Carvalho}, for which we considered a threshold velocity $v'$ = 5 $\times$ 10$^{-4}$ m/s for the transition between static and dynamic conditions. Sensitivity tests of the used coefficients are available in Carvalho et al. \cite{Carvalho}. In our simulations, we adopted a value of $E$ for the steel two orders of magnitude smaller than the real one ($E$ $=$ 1.96 $\times$ 10$^{11}$ Pa), which allows increasing the time step while keeping a reasonable accuracy in the results \cite{Lommen}. The mechanical properties of the materials used are listed in Tabs. \ref{tabmaterials} and \ref{tabcoefficients}.

\begin{table}[!h]
	\centering
	\caption{Properties of materials used in the simulations: $E$ is the Young's modulus, $\nu$ is the Poisson ratio, $\rho$ is the material density, and $d$ is the particle diameter.}
	\label{tabmaterials}
	\begin{tabular}{l|c|c|c|c|c}
		\hline
		& \textbf{Material} & \textbf{$E$ (Pa)} & \textbf{$\nu$} & \textbf{$\rho$ (kg/m$^{3}$)}& \textbf{$d$ (mm)}\\
		\hline
		Intruder & Steel\footnotesize{$^{(1)}$} & $1.96 \times 10^{9}$  & 0.29 & 7800 & $d_{int} =$ 16            \\
		Grains & Polyurethane\footnotesize{$^{(1),(2)}$} & $4.14 \times 10^{6}$   & 0.50 & 1280 & $d_{s}$ = 4; $d_{l}$ = 5           \\
		Walls & Glass\footnotesize{$^{(1)}$} & $0.64 \times 10^{11}$ & 0.23 & 2500 & $L_x$ = 400; $L_{y}$ = 400\\    
		\hline
		\multicolumn{3}{l}{\footnotesize{$^{(1)}$ Hashemnia and Spelt \cite{Hashemnia}}}\\
		\multicolumn{3}{l}{\footnotesize{$^{(2)}$ Gloss \cite{Gloss}}}
	\end{tabular}
\end{table}

\begin{table}[!h]
	\centering
	\caption{Coefficients and threshold used in the numerical simulations.}
	\label{tabcoefficients}
	\begin{tabular}{l|c|c}
		\hline
		\textbf{Coefficient}  & \textbf{Symbol} & \textbf{Value} \\
		\hline		  
		Restitution coefficient (grain-grain) & $\epsilon_{gg}$ & 0.3 \\
		Restitution coefficient (grain-intruder)\footnotesize{$^{(2)}$} & $\epsilon_{gi}$ & 0.7 \\
		Restitution coefficient (grain-wall)\footnotesize{$^{(3)}$} & $\epsilon_{gw}$ & 0.7 \\
		Dynamic friction coefficient (grain-grain)\footnotesize{$^{(1)}$} & $\mu_{gg}$ & 1.2 \\
		Dynamic friction coefficient (grain-intruder)\footnotesize{$^{(2)}$} & $\mu_{gi}$ & 1.8 \\
		Dynamic friction coefficient (intruder-bottom wall) & $\mu_{iw}$ & 0.7 \\  
		Dynamic friction coefficient (grain-walls)\footnotesize{$^{(1)}$} & $\mu_{gw}$ & 0.4 \\		
		Static friction coefficient (grain-bottom wall) & $\mu_{s,gw}$ & 0.7 \\
		Threshold velocity (dynamic/static friction) & $v^{\prime}$ & $v'$ = 5 $\times$ 10$^{-4}$ m/s\\
		\hline
		\multicolumn{3}{l}{\footnotesize{$^{(1)}$ Carvelaro et al. \cite{Carlevaro}}} \\
		\multicolumn{3}{l}{\footnotesize{$^{(2)}$ Hashemnia et al. \cite{Hashemnia}}} \\
		\multicolumn{3}{l}{\footnotesize{$^{(3)}$ Gondret et al. \cite{Gondret}}}
	\end{tabular}
\end{table}

\begin{table}[!h]
	\centering
	\caption{Packing fraction $\phi$, and numbers of small $N_s$ and large $N_l$ grains  used in the simulations}
	\label{tab_number_grains}
	\begin{tabular}{|c|c|c|}
		\hline
		$\phi$ & $N_s$ & $N_l$\\
		\hline
		0.76  & 4832 & 3092 \\ \hline
		0.79  & 5023 & 3215 \\ \hline
		0.80  & 5087 & 3256 \\ \hline
		0.805 & 5118 & 3276 \\ \hline
		0.806 & 5125 & 3280 \\ \hline
		0.807 & 5131 & 3284 \\ \hline
		0.81  & 5150 & 3296 \\ \hline		
	\end{tabular}
\end{table}

\begin{figure}[ht]
	\centering
	\includegraphics[width=0.6\columnwidth]{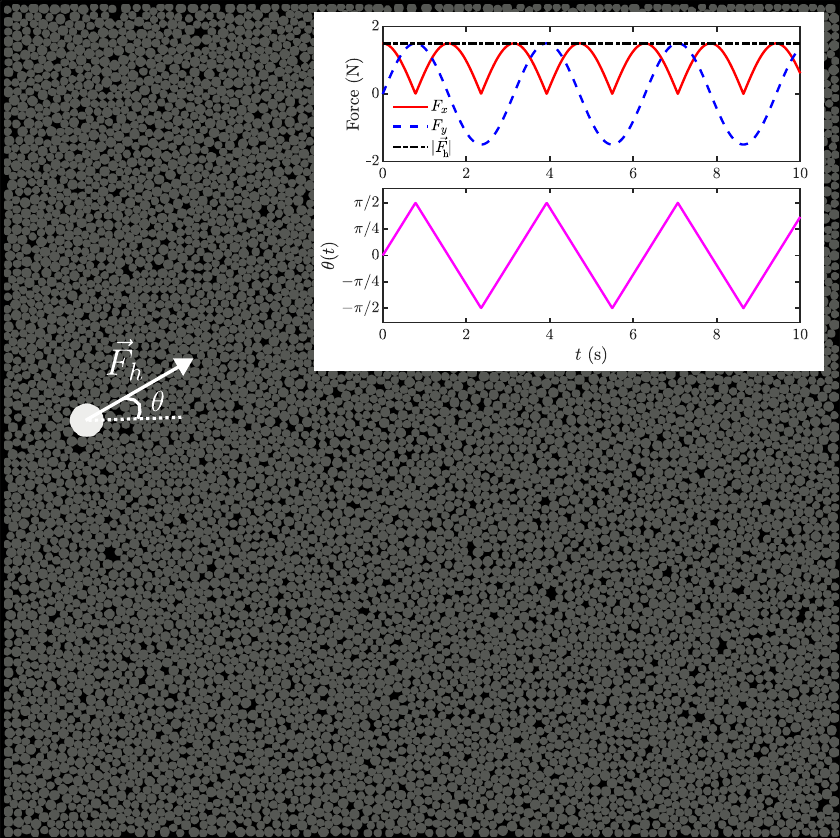}
	\caption{Layout of the numerical setup. The horizontal plane is shown in black, the disks in gray, and the intruder in white color. The figure also shows the direction of application of the horizontal force at a given time instant. The inset shows one example of time evolution of the applied force $\vec{F_h}$: on the top, the values of the $x$ and $y$ components of the force, $F_x$ and $F_x$, respectively, and we can observe that $|\vec{F_h}|$ remains constant; on the bottom, the angle $\theta$ with respect to the $x$ coordinate.}
	\label{fig:setup}
\end{figure}

We forced the intruder to move within the disks under the influence of an external force $\vec{F}_h(t)$, given by Eq. \ref{eqHarmonicforce}, which oscillates in direction within the $xy$ plane. We do not consider any motion in the direction perpendicular to the $xy$ plane, ensuring that no collisions occur between the disks and the bottom wall. The intruder is initially placed at the position $x_i$ $=$ -160 mm, $y_i$ $=$ 0 mm, on the left side of the simulation cell, and is driven from left to right through the granular medium toward its final position at $x_f$ $=$ 160 mm (and a coordinate $y_f$, which depends on the system conditions). Thus, for all values of $F_0$, $\omega$, and $\phi$ used in the simulations, the intruder was intended to traverse a total longitudinal distance of $\Delta x = 320$~mm. However, due to the nature of the applied forcing and the non-linear resistance encountered within the medium, this was not always achieved. Figure \ref{fig:setup} shows a layout of the numerical setup, and Fig. \ref{fig:trajectories} illustrates examples of the trajectories followed by the intruder. Files with the numerical setup of our simulations, output files, and scripts for post-processing the outputs are available in an open repository \cite{Supplemental2}.

\section{\label{sec:Res} RESULTS AND DISCUSSION}

\subsection{\label{sec:trajectories} Intruder's trajectories and velocities}

\begin{figure}[ht]
	\centering
	\includegraphics[width=0.75\columnwidth]{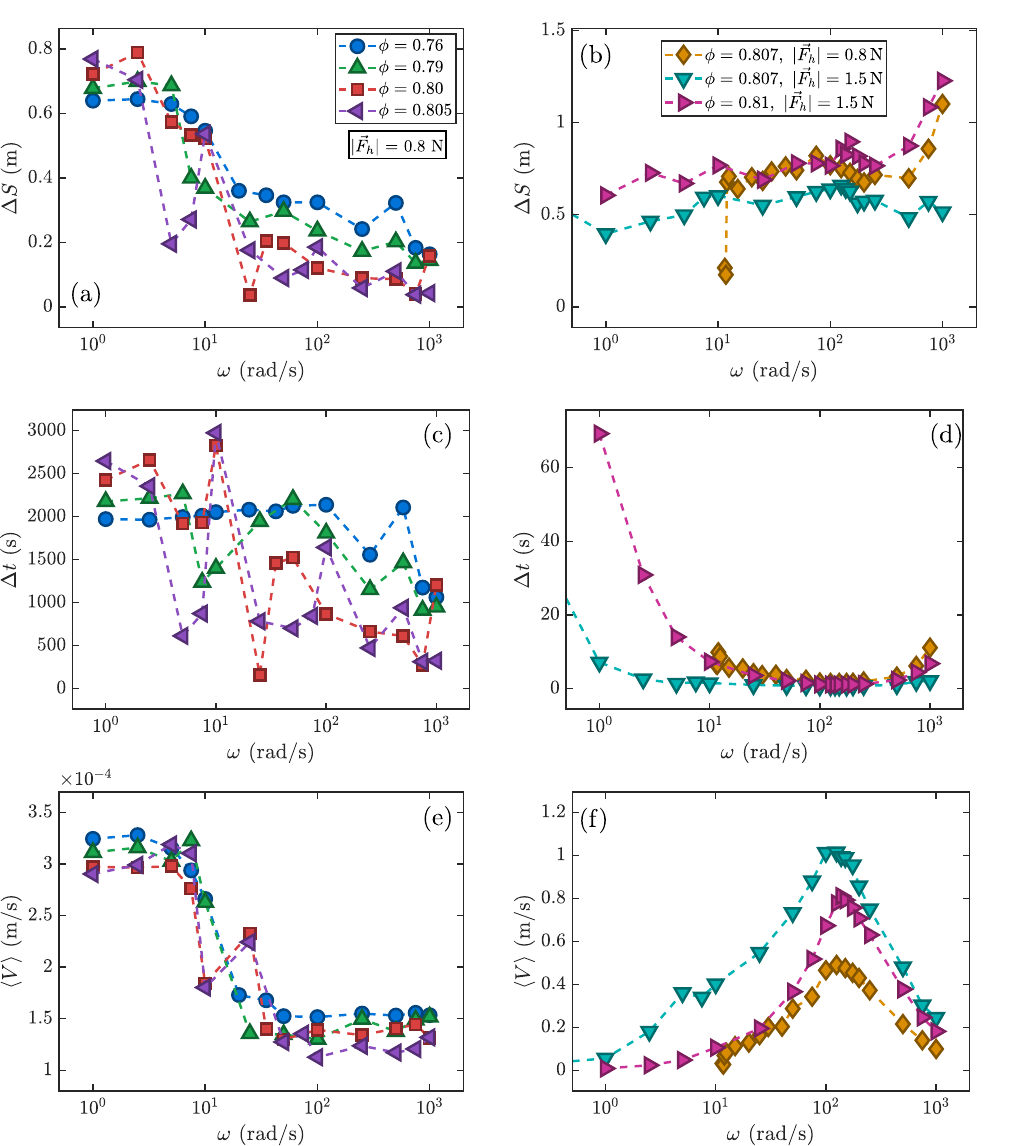}
	\caption{(a-b) Total displacement $\Delta S$ of the intruder as a function of the angular frequency $\omega$: (a) low packing fractions; (b) high packing fractions. (c-d) Time duration $\Delta t$ of displacement as a function of the angular frequency $\omega$: (c) low packing fractions; (d) high packing fractions. (e-f) Average velocity $\left< V \right>$ of the intruder as a function of the angular frequency $\omega$: (e) low packing fractions; (f) high packing fractions. The corresponding values of $\phi$ are shown in the key of panels (a) and (b).}
	\label{displacements}
\end{figure}

We begin by looking into the displacement of the intruder under different packing fractions $\phi$ and angular frequencies $\omega$. For each test, we computed the total distance $\Delta S$ traveled by the intruder and the time interval $\Delta t$ it took to travel this distance. The results are shown in Figs. \ref{displacements}(a) and \ref{displacements}(b) for $\Delta S$ for low ($\phi$ $<$ 0.806) and high ($\phi$ $>$ 0.806) packing fractions, respectively, and in Figs.  \ref{displacements}(c) and \ref{displacements}(d) for $\Delta t$ for low ($\phi$ $<$ 0.806) and high ($\phi$ $>$ 0.806) packing fractions, respectively. We observe that values of $\Delta S$ are of the same order of magnitude for both low and high packing fractions, the main difference being that $\Delta S$ decreases strongly between $\omega$ $=$ 10 and 10$^{2}$ rad/s for $\phi$ $<$ 0.806, while it increases slowly with $\omega$ for $\phi$ $>$ 0.806. \corr{Please note that in some cases part of the system (in front of the intruder) jams, so that the intruder stops and the total distance traveled is smaller than the system length.} On the other hand, the time duration $\Delta t$ is much higher when $\phi$ $<$ 0.806 (values of up to four orders of magnitude higher) than when $\phi$ $>$ 0.806. As we will show next, below a given threshold for compaction (corresponding to $\phi$ $=$ 0.806) the grains are more free to slide over the base and dissipate much more energy by basal friction than when above the threshold. This, added to the fact that dissipation by basal friction is higher than that by damping due to periodic deformations of contact chains (that is the mechanism dominant when $\phi$ $>$ 0.806), makes the intruder take more time to complete its motion when under small packing fractions.

To estimate the speed of displacement, we computed an averaged velocity $\left< V \right>$ as,

\begin{equation}
	\left< V \right> = \frac{\Delta S}{\Delta t} \,\,.
\end{equation}

As a reflection of larger displacement times for small packing fractions, we expect the existence of two distinct behaviors: for $\phi$ $<$ 0.806, $\left< V \right>$ has very small values and a monotonic variation with $\omega$; for $\phi$ $>$ 0.806, $\left< V \right>$ has much higher values and a non-monotonic variation with $\omega$. Figures \ref{displacements}(e) and \ref{displacements}(f) show $\left< V \right>$ as a function of the angular frequency $\omega$ for relatively low ($\phi$ $<$ 0.806) and high ($\phi$ $>$ 0.806) packing fractions, respectively. We, indeed, observe that under low packing fractions the intruder has small effective velocities, with $\left< V \right>$ decreasing monotonically from roughly 3.0 $\times$ 10$^{-4}$ m/s when $\omega$ $\leq$ 10 rad/s to approximately 1.5 $\times$ 10$^{-4}$ m/s when $\omega$ $=$ 10$^{3}$ rad/s. For high packing fractions, Fig. \ref{displacements}(f) shows that velocities are relatively small when $\omega$ $\leq$ 1 to 10 rad/s, with $\left< V \right>$ $\sim$ 10$^{-3}$ to 10$^{-2}$ m/s for the cases in shown in the figure, while values of up to three orders of magnitude higher are reached when $\omega$ $\sim$ 10$^{2}$ rad/s. The variation of $\left< V \right>$ with $\omega$ is non-monotonic, $\left< V \right>$ increasing for $\omega$ $=$ 1 rad/s on, reaching a peak at $\omega$ $\sim$ 10$^{2}$ rad/s, and decreasing toward much smaller values as $\omega$ increases further. While the small velocities at low packing fractions are explained by higher dissipation due to basal friction, the non-monotonic behavior for higher packing fractions can be explained by a resonance mechanism, as shown in Subsection \ref{sec:minimum_model}.

\begin{figure}[ht]
	\centering
	\includegraphics[width=0.85\columnwidth]{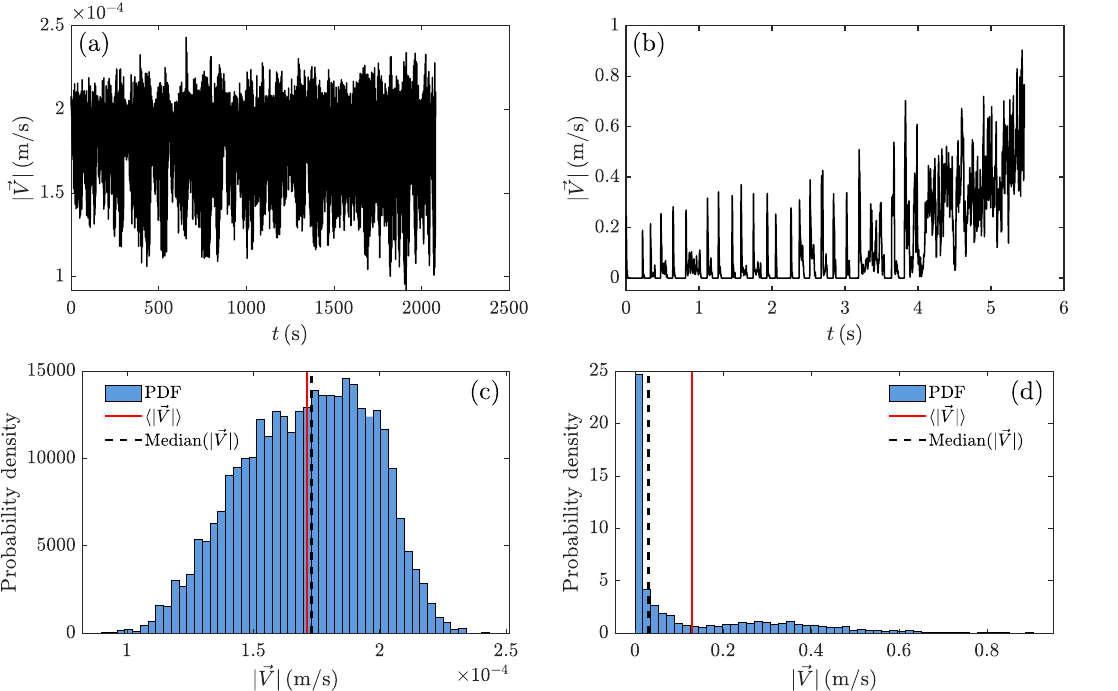}
	\caption{(a-b) Absolute values of the instantaneous velocity, $|\vec{V}|$, for $|\vec{F}_h|$ $=$ 0.8 N, $\omega$ $=$ 20 rad/s, and (a) $\phi$ $=$ 0.760 and (b) 0.807. (c-d) Histograms of $|\vec{V}|$ for $|\vec{F}_h|$ $=$ 0.8 N, $\omega$ $=$ 20 rad/s, and (c) $\phi$ $=$ 0.760 and (d) 0.807.}
	\label{fig:velocities}
\end{figure}

We computed also the instantaneous velocity of the intruder,

\begin{equation}
	\vec{V} = \frac{d \vec{S}}{d t} \,\,,
\end{equation}

\noindent where $\vec{S}$ $=$ $x\hat{i}$ $+$  $y\hat{j}$. Figures \ref{fig:velocities}(a) and \ref{fig:velocities}(b) show two examples of the time evolution of the absolute values of the instantaneous velocity, $|\vec{V}|$, for $|\vec{F}_h|$ $=$ 0.8 N, $\omega$ $=$ 20 rad/s, and, respectively, $\phi$ $=$ 0.760 and 0.807. We observe that under low packing fractions the intruder presents relatively low values of velocities that fluctuate under a roughly constant mean value throughout its motion. Those fluctuations are mostly due to the formation and breakage of contact chains \corr{(Carvalho et al. \cite{Carvalho} showed that grains within load-bearing chains undergo creep during chain rupture, leading eventually to chain collapse and enabling the intruder to progress in an intermittent way)}. On the other hand, when under high packing fractions the intruder has a highly intermittent motion, with long periods of very low velocities (or even no motion), and small periods in which it presents a high peak in the velocity (three to four orders of magnitude higher). This indicates that a great portion of the imposed force $\vec{F}_h$ is stored in the deformation of contact chains, and suddenly released by when its direction changes. These behaviors for low and high packing fractions are corroborated by the histograms of $|\vec{V}|$ shown in Figs. \ref{fig:velocities}(c) and \ref{fig:velocities}(d), in which we notice a smooth distribution around the mean value for $\phi$ $=$ 0.760 and a large concentration close to zero with a long tail of much higher velocities when $\phi$ $=$ 0.807. The dynamics of these behaviors can be observed in the movies available in the supplemental material \cite{Supplemental}.

\begin{figure}[ht]
	\centering
	\includegraphics[width=0.85\columnwidth]{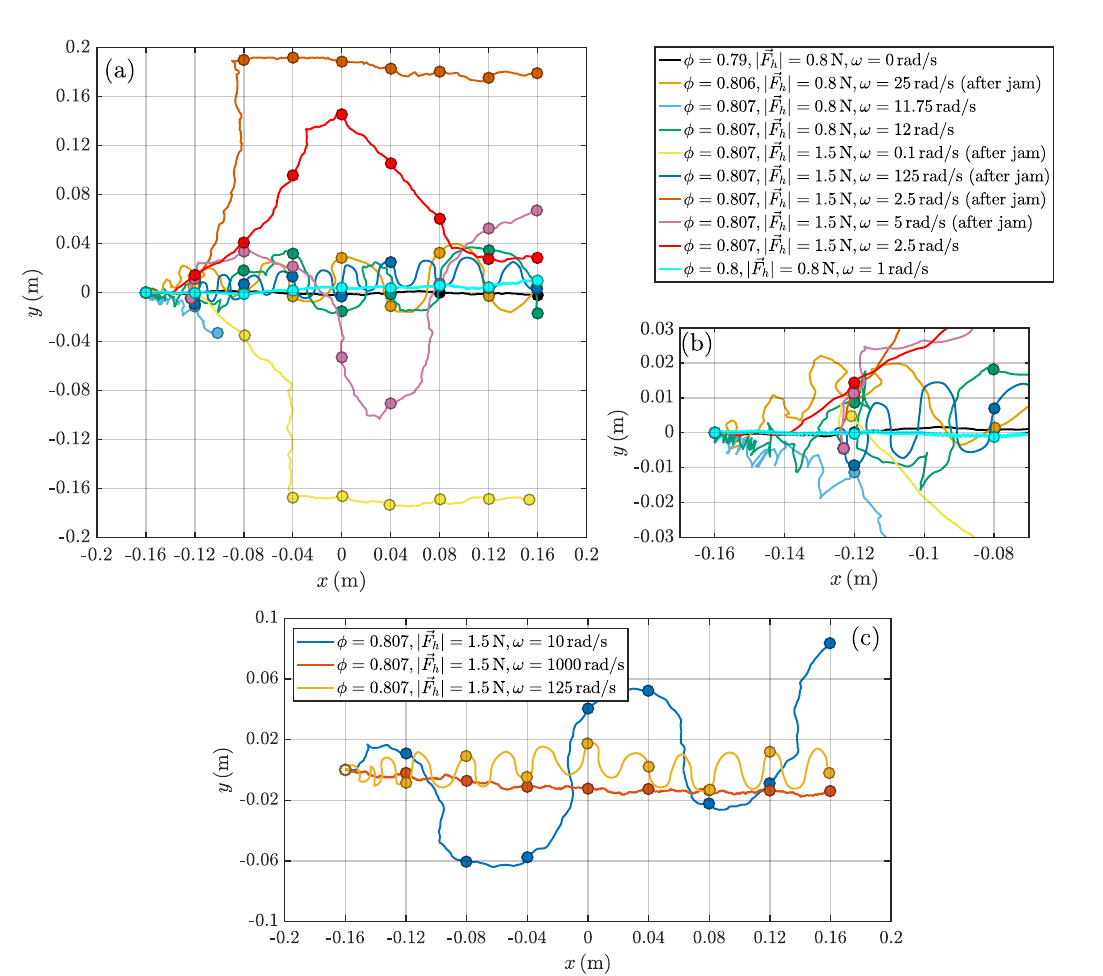}
	\caption{(a) Intruder trajectories for different values of $\phi$, $\omega$, and $|\vec{F}_h|$. The legends are in the figure key. (b) Zoom in the region within approximately -0.16 and -0.08 m (beginning of the motion). (c) Trajectories for $|\vec{F}_h|$ $=$ 1.5 N, $\phi$ $=$ 0.807, and $\omega$ $=$ 10, 125, and 1000 rad/s. The legend is in the figure key.}
	\label{fig:trajectories}
\end{figure}

In order to further investigate how the intruder displacement occurs for the different forcings and packing fractions, we show in Fig. \ref{fig:trajectories} the trajectories of the intruder for different values of $\phi$, $\omega$, and $|\vec{F}_h|$. Figure \ref{fig:trajectories}(a) shows the trajectories over the entire domain, Fig. \ref{fig:trajectories}(b) is a zoom in the region within approximately -0.16 and -0.08 m (showing the beginning of the motion), and Fig. \ref{fig:trajectories}(c) compares the trajectories for three different values of $\omega$ when $|\vec{F}_h|$ $=$ 1.5 N and $\phi$ $=$ 0.807, and the legends are shown the figure keys. In Figs. \ref{fig:trajectories}(a) and \ref{fig:trajectories}(b), there are some cases in which ``after jam'' is indicated. In those cases, we initially imposed a forcing that did not oscillate ($\omega$ = 0 rad/s), so that the intruder moved until the system jammed. Once jamming occurred, a harmonic force was applied (with the frequencies indicated in the figure key) to investigate whether it could induce motion. As Figs. \ref{fig:trajectories}(a) and \ref{fig:trajectories}(b) show, the motion resumed after applying an oscillatory forcing, which represents, therefore, an effective mechanism for generating movement under such circumstances. This observation is corroborated, for instance, by the case $\phi$ $=$ 0.807 and $|\vec{F}_h|$ $=$ 0.8 N shown in Figs. \ref{displacements}(b), \ref{displacements}(d) and \ref{displacements}(f), for which motion occurs only when $\omega$ exceeds a certain threshold; below this value, no motion is observed, and the system remains jammed.

For $\omega$ $\leq$ 125 rad/s, we basically observe that when $\phi$ $<$ 0.806 the intruder moves mainly in the longitudinal direction, and the probability of jamming is low. On the other hand, when $\phi$ $\geq$ 0.806 the intruder experiences a considerable transverse displacement, following linear paths that change direction when $\omega$ is small ($\omega$ $\lesssim$ 2.5 rad/s) and an oscillatory motion that is roughly sinusoidal when $\omega$ is relatively large ($\omega$ $\gtrsim$ 2.5 rad/s). The wavelength of trajectories increases and their amplitude decreases with  $\omega$. In addition, the region in front of the intruder sometimes jams when $\phi$ $\geq$ 0.806. These behaviors are consistent with the fact that the disks can slide and a void region opens in front of the intruder as it moves under low packing fractions, so that it can follow a linear path, while under high packing fractions the intruder moves intermittently toward regions of lower packing fractions when the applied and stored forces allow. For $\omega$ $>$ 125 rad/s, Fig. \ref{fig:trajectories}(c) shows that high $\omega$ (in this case 1000 rad/s) decreases so much the oscillating amplitude that the intruder trajectory becomes almost linear. Examples of trajectories can be observed in the movies available in the supplemental material \cite{Supplemental}.

\begin{figure}[ht]
	\centering
	\includegraphics[width=\columnwidth]{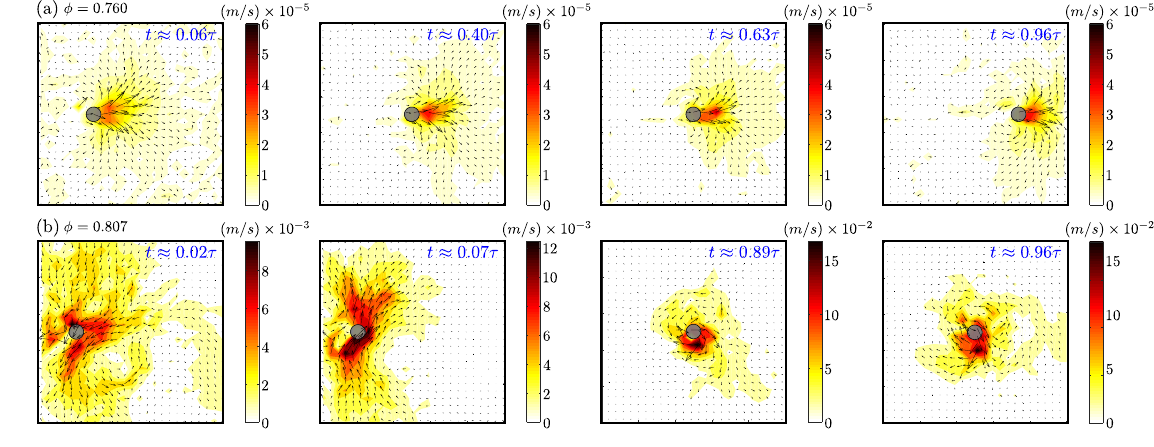}
	\caption{\corr{Snapshots of velocity fields at different time instants, indicated in each panel. The time $t$ is shown as percentages of the total time $\tau$, and the cases correspond to those shown in Fig. \ref{fig:velocities}, namely $|\vec{F}_h|$ $=$ 0.8 N, $\omega$ $=$ 20 rad/s, and (a) $\phi$ $=$ 0.760 and (b) 0.807.}}
	\label{fig:velocity_fields}
\end{figure}

\corr{Finally, Figs. \ref{fig:velocity_fields}(a) and \ref{fig:velocity_fields}(b) show snapshots of the velocity fields of the granular medium at different time instants, for $\phi$ $=$ 0.760 and $\phi$ $=$ 0.807, respectively. The corresponding time instant $t$ is indicated in each panel as a percentage of the total time $\tau$ of motion. For $\phi$ $=$ 0.760, we observe that the grains close to the intruder are pushed forward in the region in front of the intruder, and that the grains in other regions recirculate around it. This is in agreement with the fields reported in Seguin et al. \cite{Seguin1}. In this picture, the disks slide considerable distances, engendering a relatively large dissipation by basal friction (as shown in Subsections \ref{sec:contacts} and \ref{sec:optimal_frequencies}). For $\phi$ $=$ 0.807, on the other hand, the higher packing fraction makes grains compress and decompress at some instants (along contact chains), depending on the direction of the applied force (as shown in Subsection \ref{sec:contacts}), and to slide at other instants (when local packing fraction allows). This is indeed what is shown in Fig. \ref{fig:velocity_fields}(b), in which the grains close to the intruder decompress at $t$ $\approx$ 0.02$\tau$ (toward the left bottom in the panel) and $t$ $\approx$ 0.07$\tau$ (toward the up right in the panel), and slide around the obstacle at $t$ $\approx$ 0.89$\tau$ and $t$ $\approx$ 0.96$\tau$.}

\subsection{\label{sec:force_levels} Force levels}

\begin{figure}[ht]
	\centering
	\includegraphics[width=0.85\columnwidth]{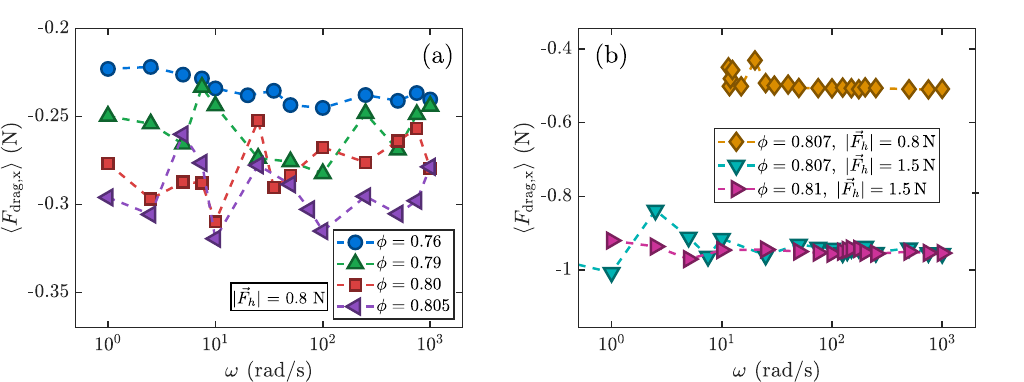}
	\caption{Drag force in the longitudinal direction as a function of $\omega$: (a) low packing fractions; (b) high packing fractions.}
	\label{fig:forces}
\end{figure}

We computed the drag force acting on the intruder, and then time averaged its longitudinal component, obtaining $\left< F_{drag,x} \right>$. Figures \ref{fig:forces}(a) and \ref{fig:forces}(b) show $\left< F_{drag,x} \right>$ as a function of $\omega$ for low and high packing fractions, respectively. We observe that although $\left< V \right>$ varies with $\omega$ (and also the dissipated power, as shown next in Subsection \ref{sec:optimal_frequencies}), values of $\left< F_{drag,x} \right>$ remain roughly constant along the different values of $\omega$ simulated. This can be seen as a time-averaged behavior of the quasistatic regime, where drag is given by friction and is independent of the velocity \cite{Andreotti_6, Seguin1, Carvalho}. With respect to $\phi$, we observe that the modulus of $\left< F_{drag,x} \right>$ increases with the packing fraction as a consequence of the higher confinement.

\subsection{\label{sec:contacts} Solid-solid contacts}

\begin{figure}[ht]
	\centering
	\includegraphics[width=0.95\columnwidth]{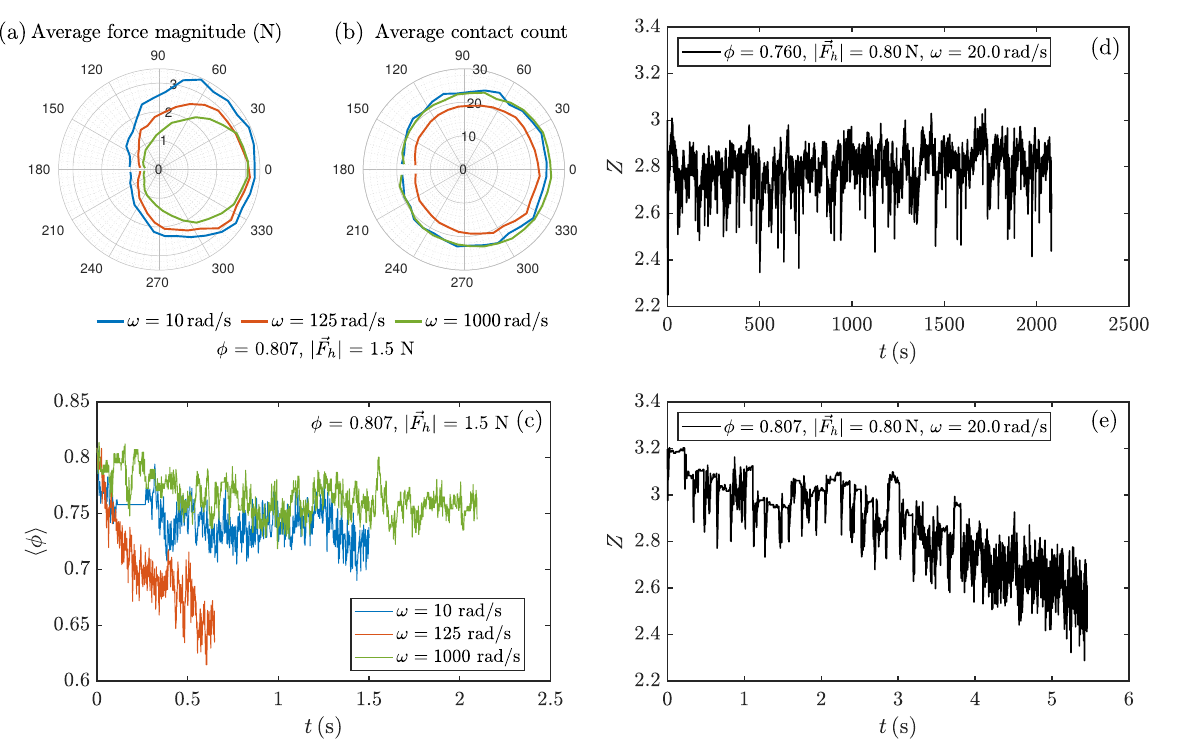}
	\caption{Number of contacts and transmitted forces for an intruder moving when $\phi$ $=$ 0.760 and 0.807. (a) Time average of summed forces at grain-grain contacts close to the intruder, according to the angle with respect to the longitudinal direction (0$^{\circ}$), for $\vec{F}_h$ $=$ 1.5 N, $\phi$ $=$ 0.807, and $\omega$ $=$ 10, 125, and 1000 rad/s. The time-averaged summed force was computed by summing the magnitudes of grain-grain contact forces, for each angle, in a region with radius equal to 2.5$d_{int}$ around the intruder, and the results were afterward averaged in time. (b) Average number of the total grain-grain contacts close to the intruder (for the cases of panel (a)). (c) Time evolution of the average packing fraction $\left< \phi \right>$ in the region close to the intruder. Panels (a-c) are parameterized by $\omega$, whose values are listed in the key of panel (c), and the system packing fraction was $\phi$ $=$ 0.807. (d-e) Example of time evolution of the coordination number $Z$ computed in the a region with radius equal to 2$d_{int}$ around the intruder when $\vec{F}_h$ $=$ 0.8 N, $\omega$ $=$ 20 rad/s, and $\phi$ $=$ 0.760 and 0.807, respectively.}
	\label{fig:contacts}
\end{figure}

In order to better understand how the intruder moves when under high packing fractions, we computed the number of contacts and the corresponding transmitted force in a region close to the intruder, for $\phi$ $=$ 0.760 and 0.807. Figure \ref{fig:contacts}(a) shows the time-averaged summed force felt by the grains close to the intruder, according to the angle with respect to the longitudinal direction (0$^{\circ}$). The diagram is in polar coordinates, and was computed for $\vec{F}_h$ $=$ 1.5 N, $\phi$ $=$ 0.807, and $\omega$ $=$ 10, 125, and 1000 rad/s. The summed force was computed by summing, for each angle around the center of the intruder, the magnitude of the grain-grain contact forces in a region with radius equal to 2.5$d_{int}$ around the intruder, for each stored time (corresponding to every 375 time steps, i.e., every 0.0012 s). The time-averaged summed force was then obtained by averaging the results in time. The same procedure was followed for the number of solid-solid contacts and the average packing fraction around the intruder. We notice in Fig. \ref{fig:contacts}(a) that, although $\omega$ $=$ 125 rad/s engenders the highest displacement velocities, the average force is within the values found for the other angular velocities.  We also observe a clear asymmetry with respect to the angle, with much larger force levels in the region in front of the intruder than behind it. For the same angular velocities, Fig. \ref{fig:contacts}(b) shows that the average number of contacts close and around the intruder is lesser for the optimum angular velocity, $\omega$ $=$ 125 rad/s. This indicates that in this case the intruder has a higher probability of finding space to move while the direction of the applied force oscillates. Considering the average packing fraction $\left< \phi \right>$ in the region close and around the intruder, Fig. \ref{fig:contacts}(c) shows that it decreases slowly with time for $\omega$ $=$ 10 and 1000 rad/s, and much faster when $\omega$ $=$ 125 rad/s. This corroborates the  higher probability of finding more space to move when $\omega$ $=$ 125 rad/s. \corr{Figures \ref{fig:contacts}(d) and \ref{fig:contacts}(e)} show the coordination number $Z$ computed in a region with radius equal to 2$d_{int}$ around the intruder when $\vec{F}_h$ $=$ 0.8 N, $\omega$ $=$ 20 rad/s, and, respectively, $\phi$ $=$ 0.760 and 0.807. We observe that the average value of $Z$ remains roughly constant when $\phi$ $=$ 0.760 and decreases along time when $\phi$ $=$ 0.807. For the latter case, we observe in Fig. \ref{fig:velocities}(b) an increase in the intruder's velocity in the region corresponding to the decrease of $Z$.

\begin{figure}[ht]
	\centering
	\includegraphics[width=\columnwidth]{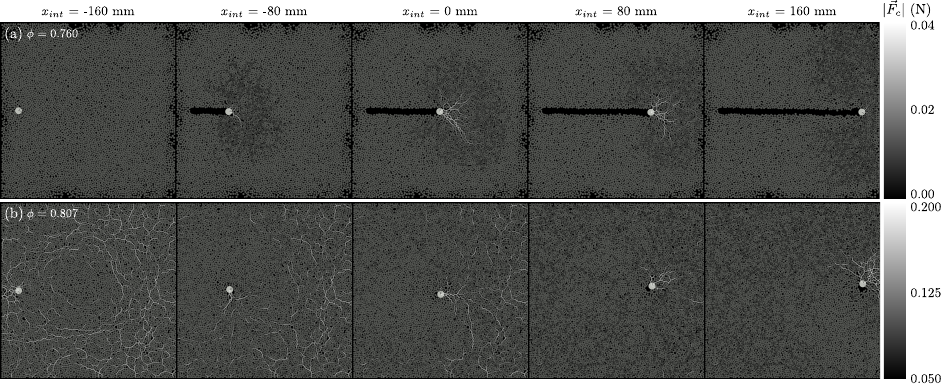}
	\caption{\corr{Snapshots of load-bearing force chains for different intruder positions, indicated above each panel. The cases correspond to those shown in Fig. \ref{fig:velocities}, namely $|\vec{F}_h|$ $=$ 0.8 N, $\omega$ $=$ 20 rad/s, and (a) $\phi$ $=$ 0.760 and (b) 0.807. The intensity of transmitted load is shown in gray scale (from dark to clear gray as the transmitted load increases).}}
	\label{fig:contact_network}
\end{figure}

\corr{Finally, Figs. \ref{fig:contact_network}(a) and \ref{fig:contact_network}(b) show snapshots with the load-bearing force chains for different intruder positions (indicated in each panel), for $\phi$ $=$ 0.760 and (b) 0.807, respectively. We observe clearly longer and dense contact chains bearing high load in the $\phi$ $=$ 0.807 case. This means that a larger fraction (comparatively to the $\phi$ $=$ 0.760 case) of the applied force is then stored elastically in the deformation of the contact network, especially along long force chains.}

\subsection{\label{sec:optimal_frequencies} Power consumption, packing fraction and optimal frequencies}

\begin{figure}[ht]
	\centering
	\includegraphics[width=0.75\columnwidth]{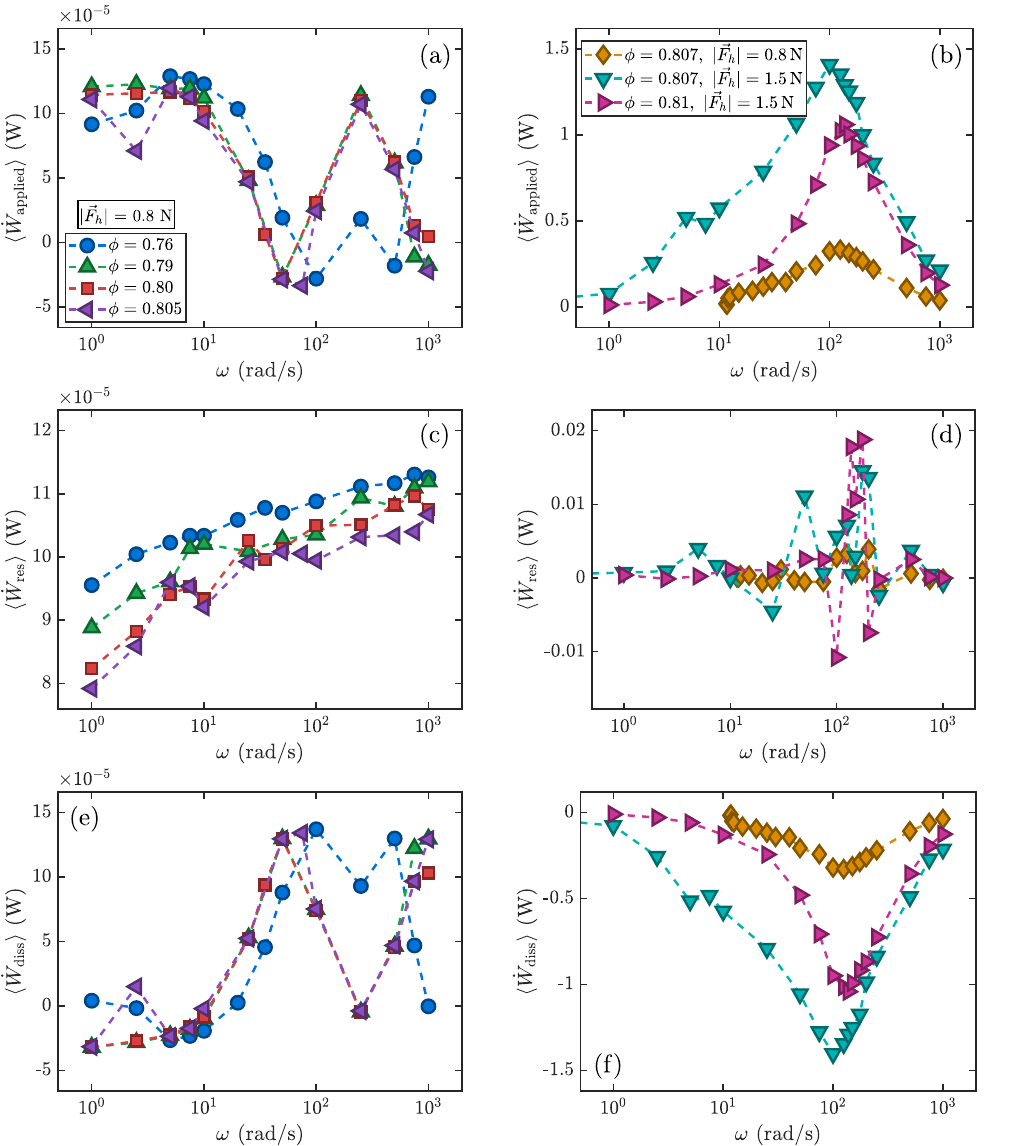}
	\caption{(a-b) Applied power $\left< \dot{W}_{applied} \right>$ used for the intruder's motion as a function of the angular frequency $\omega$: (a) low packing fractions; (b) high packing fractions (a). (c-d) Power due to the resultant force, $\left< \dot{W}_{res} \right>$, as a function of $\omega$: (c) low packing fractions; (d) high packing fractions. (e-f) Dissipated power $\left< \dot{W}_{diss} \right>$ as a function of $\omega$: (e) low packing fractions; (f) high packing fractions.}
	\label{fig:power}
\end{figure}

In order to evaluate how the efficiency of the intruder's motion vary with the oscillation frequency, we computed the applied power for displacing the intruder as

\begin{equation}
	\dot{W}_{applied} = \vec{F}_h \cdot \vec{u}_{int} \,\,,
\end{equation}  

\noindent where $\vec{u}_{int}$ is the intruder velocity, and we also computed the power $\dot{W}_{res}$ due to the resultant force $\vec{F}_{res}$ acting on the intruder,

\begin{equation}
	\dot{W}_{res} = \vec{F}_{res} \cdot \vec{u}_{int} \,\,.
\end{equation} 

\noindent With these two powers, we estimated the dissipated power as

\begin{equation}
	\dot{W}_{diss} = \dot{W}_{res} - \dot{W}_{applied} \,\,.
\end{equation}

Figures \ref{fig:power}(a) and \ref{fig:power}(b) show the average of the applied power $\left< \dot{W}_{applied} \right>$ for low and high packing fractions, respectively, as a function of the angular frequency $\omega$. For the low packing fractions ($\phi$ $<$ 0.806), we observe that higher values occur for $\omega$ $\leq$ 10 rad/s, with a decrease in $\left< \dot{W}_{applied} \right>$ as frequencies are increased. However, a peak appears within 200 rad/s $\leq$ $\omega$  $\leq$ 300 rad/s. For the high packing fractions ($\phi$ $>$ 0.806), $\left< \dot{W}_{applied} \right>$ is low when $\omega$ $\leq$ 10 rad/s, increasing and then decreasing strongly as $\omega$ increases, with a clear peak at $\omega$ $\approx$ 10$^{2}$ rad/s. On the other hand, Figures \ref{fig:power}(c) and \ref{fig:power}(d) show relatively smoother variations of mean averages of the power due to the resultant force $\left< \dot{W}_{res} \right>$. In the case of high packing fractions, we observe, however, peaks and depressions when $\omega$ $\sim$ 10$^{2}$ rad/s. We notice also that the maxima of $\left< \dot{W}_{applied} \right>$ and $\left< \dot{W}_{res} \right>$ are much higher (by five orders of magnitude) for the high packing fractions, the reason for this being the higher velocities reached by the intruder when $\phi$ $>$ 0.806.

Figures \ref{fig:power}(e) and \ref{fig:power}(f) show the time-averaged dissipated power during the process $\left< \dot{W}_{diss} \right>$ as a function of the angular frequency $\omega$, for low and high packing fractions, respectively. For the case of high packing fractions, we observe a non-monotonic behavior, with lower values occurring for $\omega$ $\approx$ 100 rad/s. For the case of low packing fractions, the smaller values occur for low angular frequencies ($\omega$ $\leq$ 10 rad/s). These behaviors are consistent with those for $\left< V \right>$, showing that the most effective frequencies are $\omega$ $\leq$ 10 rad/s when $\phi$ $<$ 0.806 and $\omega$ $\approx$ 10$^{2}$ rad/s  when $\phi$ $>$ 0.806. We notice that a depression occurs for 200 rad/s $\leq$ $\omega$  $\leq$ 300 rad/s when $\phi$ $<$ 0.806. Therefore, although optimal frequencies in this case are equal or less than 10 rad/s, a resonant mechanism seems to occur at $\omega$ $\sim$ 10$^{2}$ rad/s, as in the case of high packing fractions. This indicates that the system has a natural frequency with this order of magnitude. This is indeed the case, as shown in Subsection \ref{sec:minimum_model}.

Based on both the displacement velocity and dissipated power, data show that the optimal motion within disks under low packing fractions ($\phi$ $<$ 0.806) occurs at angular frequencies $\omega$ $\leq$ 10 rad/s, while for high packing fractions ($\phi$ $>$ 0.806) the optimal frequency is $\omega$ $\sim$ 10$^{2}$ rad/s. In the case of the employed disks, we observe a cutoff packing fraction of $\phi$ $=$ 0.806, above which local jamming occurs more often.

\subsection{\label{sec:minimum_model} Minimum model}

We propose next a minimum model for finding the optimal frequency for displacing an intruder in a packed granular material. For that, we begin by observing that our granular system exhibits a collective elastic behavior under small strain. Inspired by the Hertz-Mindlin contact model \cite{direnzo}, we build a minimum model by approximating the granular packing as a network of force chains that acts as an effective elastic medium. As a starting point, we consider that under high packing the granular medium will behave as a simple harmonic oscillator with Hookean response. The motion of the intruder is, thus, governed by Newton’s second law with a linear restoring force \corr{and a damping term},

\begin{equation}
	\corr{m_{int}\frac{d^2x}{dt^2} + b\frac{dx}{dt} + k_{eff}x = 0 \,\,,}
	\label{eq_newton}
\end{equation}

\noindent where $m_{int}$ is the mass of the intruder\corr{, $b$ is a damping coefficient,} and $k_{eff}$ is the effective stiffness of the medium. The solution describes a harmonic motion with natural frequency given by Eq. \ref{eq_nat_freq},

\begin{equation}
	\omega_0 = \sqrt{\frac{k_{eff}}{m_{int}}} \,\,,
	\label{eq_nat_freq}
\end{equation}

\noindent \corr{and damping constant equal to $b/m_{int}$.} Equations \ref{eq_newton} and \ref{eq_nat_freq} provide the framework, but estimating $k_{eff}$ for a granular material requires deeper insight into the microstructure of the medium, particularly the network of solid-solid contact chains. For the sake of simplicity, we model the chains as assemblies of elastic grain-grain contacts, through which stresses are transmitted. Therefore, we treat each force chain as a series of elastic springs, and estimate their contribution to the system’s stiffness. By considering that a typical chain of $N_c$ disks resists the motion as springs connected in series, their total stiffness is

\begin{equation}
	\frac{1}{k_{chain}} = \sum_{i=1}^{N_c} \frac{1}{k_{contact}} \,\,,
	\label{eq_k_chain}
\end{equation}

\noindent where $k_{contact}$ is the stiffness of each contact\corrnew{. For spherical particles, the Hertzian expression reads}

\begin{equation}
	k_{contact} = \frac{4}{3} E_c (\delta_n r_c)^{1/2} \,\,.
	\label{eq_k_1_contact}
\end{equation}

\corrnew{For disk-shaped particles (case of the present work) the contact stiffness follows the formulation proposed by \citet{Herman} (details of this formulation can be found in the Supplemental Material of \citet{Herman}), given by}

\begin{equation}
	\corrnew{k_{\text{contact}} = \pi E_c h_m \, f\left( \frac{\delta_n r_c}{2 h_m^2} \right) \,\,,}
	\label{eqn_kcontact_disk}
\end{equation}

\noindent \corrnew{where $h_m$ denotes the smaller height among the interacting disks. The normal overlap $\delta_n$ is estimated from the relative radial positions of the intruder and the granular medium as}

\begin{equation}
	\corrnew{\delta_n = R_1 + R_2 - |r_1 - r_2|\,\,,}
\end{equation}

\noindent \corrnew{where $r_1 = \sqrt{x_1^2 + y_1^2}$ and $r_2 = \sqrt{x_2^2 + y_2^2}$ are the radial distances of the interacting bodies from a common reference point. Here, $R_1$ corresponds to the intruder radius, and $R_2$ is the grain radius. The quantities $r_1$ and $r_2$ are obtained directly from the DEM simulations by taking an average of instantaneous positions of the particle centers. This expression provides a simplified (radial) estimate of the interparticle distance and is used to evaluate the overlap in the contact law.}

In Eqs.\ref{eq_k_1_contact} and \ref{eqn_kcontact_disk}, $r_c$ is the effective radius,

\begin{equation}
	r_c = \frac{r_i r_j}{r_i + r_j} \,\,,
\end{equation}

\noindent \corrnew{and $E_c$} is the effective contact modulus between disks $i$ and $j$ (whose Young's moduli are $E_i$ and $E_j$, the Poisson ratios are $\nu_i$ and $ \nu_j$, and the radii are $r_i$ and $r_j$, respectively),

\begin{equation}
	\frac{1}{E_c} = \frac{1 - \nu_i^2}{E_i} + \frac{1 - \nu_j^2}{E_j} \,\,.
\end{equation}

By assuming that $k_{contact}$ is the same for all contacts, and estimating $N_c$ as the length of the chain $L_{fc}$ divided by the mean grain diameter $d_g$, Eq. \ref{eq_k_chain} becomes

\begin{equation}
	k_{chain} = \frac{k_{contact}d_g}{L_{fc}} \,\,.
	\label{eq_k_chain2}
\end{equation}

\noindent We then set $L_{fc}$ $=$ 50$d_g$, which corresponds roughly to the number of grains spanning half of the domain in the transverse direction. This choice is justified by the fact that the intruder is almost always connected to the walls through force chains.

Finally, \corr{we consider that the system consists of a certain number $N_{fc}$ of contact chains placed in parallel with one another (keeping the model simple). \corrnew{In this case, $N_{fc} = 2$ is chosen as a minimal representation of the number of dominant force chains acting in parallel, and which are in contact with the intruder. This reflects the simplified structure assumed in the model, where only a few load-bearing paths \cite{Radjai1} contribute to the effective stiffness (as seen in Fig. \ref{fig:contact_network}).} Therefore,} $k_{eff}$ is computed by \corr{considering $N_{fc}$ chains} in parallel,

\begin{equation}
	k_{eff} = N_{fc}k_{chain} = N_{fc} \frac{k_{contact}d_g}{L_{fc}} \,\,.
	\label{eqn_keff}
\end{equation}

\noindent Equation \ref{eqn_keff} can be then inserted in Eq. \ref{eq_nat_freq} for finding the natural frequency $\omega_0$ of the entire system. By inserting the values used in the simulations, we find with this simple model that $\omega_0$ $\approx$ 150 rad/s, which is close to the resonant frequency obtained from the simulations \corr{(since the damping in our system is relatively small)}. This shows that the optimal frequency for displacement of the intruder corresponds to the natural frequency of the system. If extended to a 3D system, this minimum model can indicate the optimal frequencies for perforating or moving a solid under a granular soil.

\section{\label{sec:Conclu} CONCLUSIONS}

This paper investigated numerically the motion in a granular medium of an intruder propelled by an oscillating force. We carried out DEM simulations in a quasi-2D system in which both the grains and intruder were disks, and the propeller force oscillated in direction at a given frequency. Depending on the packing fraction, oscillation frequency, and magnitude of the imposed force, we found a complex behavior for the intruder, with sideways motion, temporary blocking, and even jamming. By computing the displacement velocity, we observe a cutoff packing fraction of $\phi$ $=$ 0.806 for our system, above which local jamming occurs more often. Interestingly, we found, contrary to intuition, that in the studied system the displacement velocities are much higher (up to four orders of magnitude) when the packing fraction is above the threshold ($\phi$ $>$ 0.806). We also found that below this threshold the disks are more free to slide over the base and, consequently, dissipate much more energy by basal friction than when above the threshold. Because the dissipation by basal friction is higher than that by damping of periodic deformations of contact chains (that are the dominant mechanism when $\phi$ $>$ 0.806), the intruder takes more time to complete its motion when under small packing fractions. When $\phi$ $>$ 0.806, part of the applied force is stored in the contact chains and the intruder moves by bursts while the force direction oscillates. We computed powers based on the velocity of the intruder, engendering, therefore, dissipated powers up to five orders of magnitude higher when $\phi$ $>$ 0.806.

For our system, the results also show that under low packing fractions ($\phi$ $<$ 0.806) an optimal motion occurs at angular frequencies $\omega$ $\leq$ 10 rad/s, while for high packing fractions ($\phi$ $>$ 0.806) the optimal frequency is $\omega$ $\sim$ 10$^{2}$ rad/s. However, even for low packing fractions a depression in the dissipated power takes place for $\omega$ $\sim$ 10$^{2}$ rad/s, indicating a resonant frequency in the system. Finally, we proposed a model for the elastic response of the system that returns a natural frequency of approximately 24.5 Hz ($\omega_0$ $\approx$ 150 rad/s), explaining why a resonant mode appears at $\omega$ $\sim$ 10$^{2}$ rad/s. Our results can be further explored to find optimal methods of perforation and locomotion within sand and other grains.

\section{\label{sec:Ack} ACKNOWLEDGMENTS}

The authors are grateful to the S\~ao Paulo Research Foundation (FAPESP, Grant Nos. 2018/14981-7, 2020/04151-7 and 2024/13981-4) for the financial support provided.

\bibliography{references}

\end{document}